\documentclass[twocolumn,showpacs,preprintnumbers,amsmath,amssymb,prl]{revtex4}

\usepackage{color}
\usepackage{amsmath}
\usepackage{graphicx}
\usepackage{dcolumn}
\usepackage{bm}
\usepackage[amssymb]{SIunits}

\begin{document}

\title{Quantum gates and multi-particle entanglement by Rydberg excitation blockade and adiabatic passage}
\author{Ditte M\o ller}
\email{dittem@phys.au.dk}
\author{Lars Bojer Madsen}
\author{Klaus M\o lmer}
\affiliation{Lundbeck Foundation Theoretical Center for Quantum System Research,\\
Department of Physics and Astronomy, University of Aarhus, DK-8000, Denmark.}
\date{\today }

\begin{abstract}
We propose to apply stimulated adiabatic passage to transfer atoms from their ground
state into Rydberg excited states. Atoms a few micrometers apart experience a
dipole-dipole interaction among Rydberg states that is strong enough to shift the
atomic resonance and inhibit excitation of more than a single atom. We show that the
adiabatic passage in the presence of this interaction between two atoms leads to
robust creation of maximally entangled states and to two-bit quantum gates. For many
atoms, the excitation blockade leads to an effective implementation of
collective-spin and Jaynes-Cummings-like Hamiltonians, and we show that the adiabatic
passage can be used to generate collective $J_x=0$ eigenstates and
Greenberger-Horne-Zeilinger states of tens of atoms.
\end{abstract}

 \maketitle

Entanglement is a central property of quantum mechanical systems and an important
resource for applications in quantum information and quantum metrology. Entanglement
is subject of broad theoretical interest and experimental implementations in various
systems have been demonstrated. One scheme for creation of entanglement and quantum
gates between neutral atoms utilizes the large dipole moments of highly excited
Rydberg states resulting in a dipole-dipole interaction which is strong enough to
shift the atomic energy levels and prevent more than one atom from being excited to
the Rydberg state by a resonant laser field \cite{jaksch00,lukin01}. In alkali atoms
Rydberg states with principal quantum number $n>70$ have lifetimes $\gtrsim100$
$\mu$s and experience a dipole-dipole interaction strength, $E/\hbar$, above
$100\cdot2\pi$ MHz when atoms are separated less than $5$~$\mu$m \cite{Saffman05}.
The accompanying suppression of excitation for $n$ up to 80 \cite{Tong04,Singer04}
and the influence of the excitation blockade on coherent collective dynamics
\cite{Heidemann07} was observed in cold gases.

\begin{figure}[htbp]
  \centering
  \includegraphics[width=0.45\textwidth]{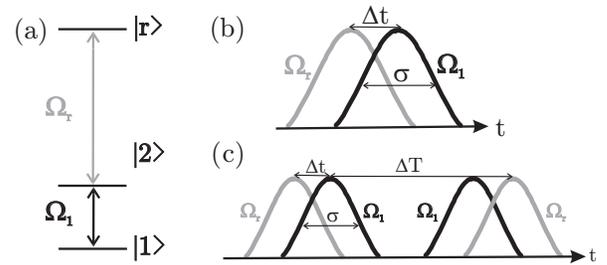}
  \caption{(a) Three-level ladder system and two laser fields with Rabi
frequencies $\Omega_1$ and $\Omega_r$. (b) STIRAP pulse sequence. The FWHM of each
pulse is $\sigma$ and the delay between pulses within one process is $\Delta t$. (c)
Pulse sequence consisting of two STIRAP processes separated by $\Delta T$ in time.}
  \label{fig:ladderandpulses}
\end{figure}

We propose to combine the Rydberg blockade mechanism with the rapid adiabatic laser
pulse sequence known from Stimulated Raman Adiabatic Passage (STIRAP)
\cite{bergmann98,gaubatz90,moller072,Deiglmayr06}. We consider atoms with two lower
levels and a \mbox{Rydberg} level in a ladder structure coupled by two resonant laser
fields with Rabi frequencies $\Omega_1$ and $\Omega_r$  as shown in
Fig.~\ref{fig:ladderandpulses}(a). The STIRAP process applies the ``counter
intuitive'' pulse sequence of Fig.~\ref{fig:ladderandpulses}(b) to transfer
population from the ground state $|1\rangle$ to a highly excited Rydberg state
$|r\rangle$ by adiabatically following \cite{messiah61} a dark state which never
populates the intermediate state $|2\rangle$. If we assume a relative phase,
$\phi_r(t)$, between the Rabi frequencies $\Omega_1$ and $\Omega_r$, a single atom
exposed to the fields follows the dark state superposition,
$|D_1\rangle=\cos\theta|1\rangle-\sin\theta{e}^{i\phi_r}|r\rangle$, where
$\tan\theta=\Omega_1/\Omega_r$ expresses the relative strength of the two laser
fields. Since the dark state has zero energy, there is no dynamic phase accumulated
during the process, but if $\phi_r$ varies, $|D_1\rangle$ acquires the geometric
phase, $\gamma_{1}=-\int \sin^2\theta d\phi_r$ \cite{berry84,moller07}. A controlled
geometric phase shift on state $|1\rangle$ can be implemented if the phase difference
between the fields evolves between the STIRAP pulses and a second inverted set of
pulses, shown in Fig.~\ref{fig:ladderandpulses}(c). For atoms with a further logical
state $|0\rangle$ such a pulse sequence enables robust geometric phase gates and
other one-bit gates \cite{moller07}.

With more atoms the Rydberg blockade comes into play, but we may still identify dark
states. For two atoms, initially in the product state $|11\rangle$ and subject to the
same interaction with the two resonant laser fields $\Omega_1$ and $\Omega_r$, the
evolution preserves the symmetry under interchange of atoms, and hence it is
sufficient to consider the Hamiltonian in the symmetric two-atomic basis
$\{|11\rangle,\frac{1}{\sqrt{2}}(|1r\rangle+|r1\rangle),\frac{1}{\sqrt{2}}(|12\rangle+|21\rangle),
|rr\rangle,\frac{1}{\sqrt{2}}(|2r\rangle+|r2\rangle),|22\rangle\}$,
\begin{small}
\begin{align}\label{eq:twoatomhamilton}
H(t)=\frac{\hbar}{2}\left[\begin{array}{cccccc}
0 & 0 & \sqrt{2}\Omega_1^* & 0 & 0 & 0\\
0 & 0 & \Omega_r^* & 0 & \Omega_1^* & 0\\
\sqrt{2}\Omega_1 & \Omega_r & 0 & 0 & 0 & \sqrt{2}\Omega_1^*\\
0 & 0 & 0 & 2E/\hbar & \sqrt{2}\Omega_r^* & 0\\
0 & \Omega_1 & 0 & \sqrt{2}\Omega_r & 0 & \sqrt{2}\Omega_r^*\\
0 & 0 & \sqrt{2}\Omega_1 & 0 & \sqrt{2}\Omega_r & 0\\
\end{array}\right],
\end{align}
\end{small}where $E$ denotes the energy shift of the state $|rr\rangle$ due to the dipole-dipole interaction. This Hamiltonian has one dark state for the
two-atom system
\begin{align}\label{eq:twoatomdark}
|D_2\rangle=&\frac{1}{\sqrt{\cos^4\theta+2\sin^4\theta}}\left[(\cos^2\theta-\sin^2\theta)|11\rangle\right.\\\notag
&\left.-\cos\theta\sin\theta{e}^{i\phi_r}(|1r\rangle+|r1\rangle)+\sin^2\theta|22\rangle\right].
\end{align}
We first assume $e^{i\phi_r}=1$ and apply the counterintuitive pulse sequence of
Fig.~\ref{fig:ladderandpulses}(b). Initially $\cos\theta=1$ and the system is in
$|D_2\rangle=|11\rangle$. Adiabaticity ensures that we remain in $|D_2\rangle$ and
after the pulses $\sin\theta=1$ and the system is in
\begin{equation}\label{eq:darkfinal}
|D_2\rangle=\frac{1}{\sqrt{2}}(-|11\rangle+|22\rangle).
\end{equation}
Equation~(\ref{eq:twoatomdark}) shows, that while the STIRAP process in the single
atom case ensures that $|2\rangle$ is never populated, due to the Rydberg blockade
the pair of atoms are adiabatically steered into a state populating $|22\rangle$.
Moreover, (\ref{eq:darkfinal}) is a maximally entangled state of the two atoms,
generated robustly irrespective of the precise pulse shapes, field strengths and the
precise value of the Rydberg interaction energy.
\begin{figure}[htbp]
  \centering
  \includegraphics[width=0.45\textwidth]{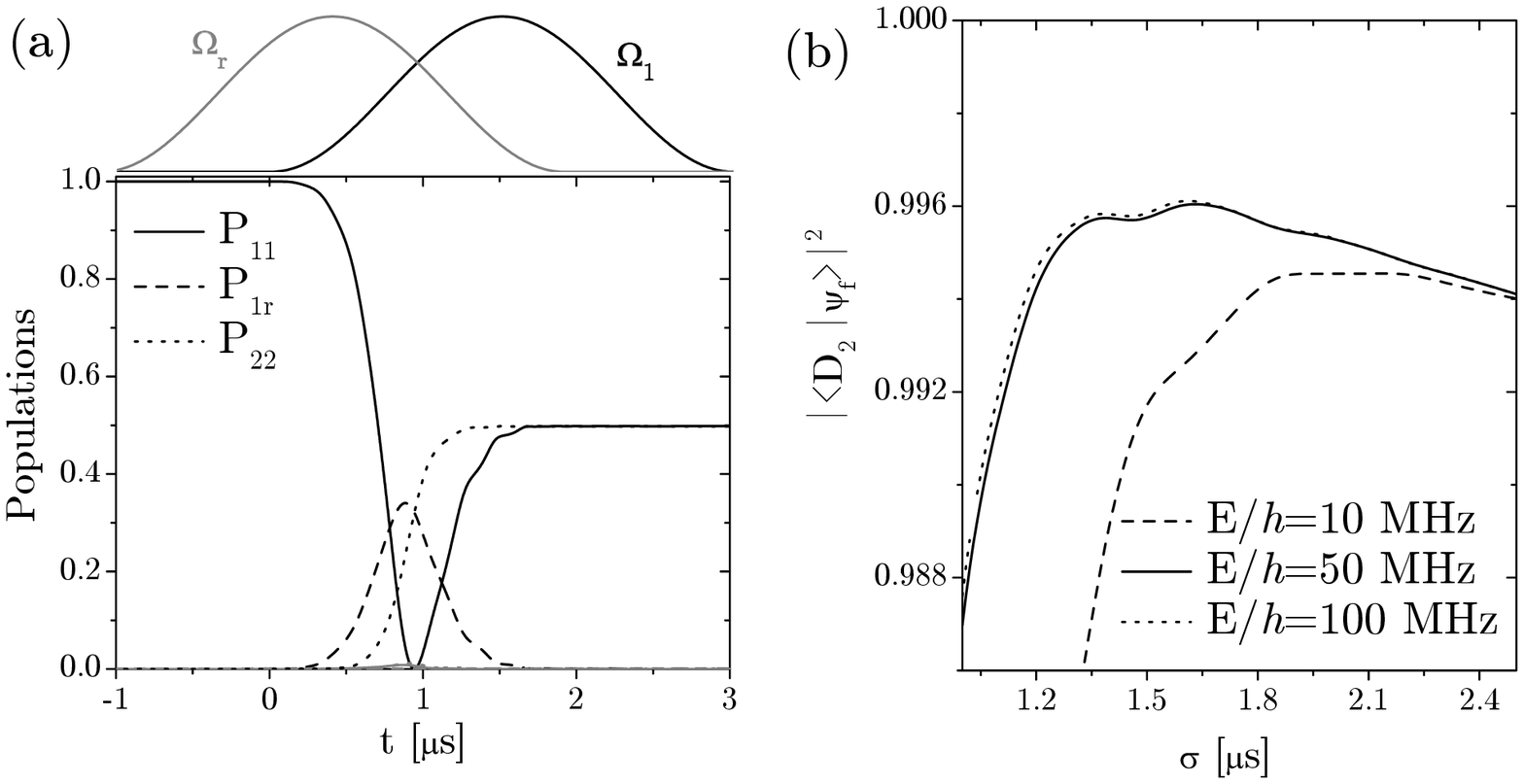}
  \caption{(a) Time evolution of the population in $|11\rangle$, $\frac{1}{\sqrt{2}}(|1r\rangle+|r1\rangle)$ and $|22\rangle$. (b) Fidelity of
the entanglement creation as a function of pulse width for different values of
dipole-dipole interaction,
  $E$. $|\psi_f\rangle$ was found by propagating the Schr\"{o}dinger
equation (\ref{eq:twoatomhamilton}) with initial state $|11\rangle$. We show the norm
square of the overlap with the target state $|D_2\rangle$. The laser pulses are
modeled by $\sin^2$-pulses, $\Omega_j(t)=\Omega_{\textrm{max},j}\sin^{2}(\frac{\pi
(t-t_{sj})}{2 \sigma})$ for $t_{sj}<t<t_{sj}+2\sigma$, where $t_{sj}$ is the instant
of time when the pulse starts and $\sigma$ the FWHM. The simulations are made with
parameters $\Omega_{max,1}/2\pi=\Omega_{max,2}/2\pi=10$ MHz, $\Delta t=1.1$ $\mu$s,
lifetime of the $|r\rangle$ state, $\tau_r=100$ $\mu$s and in (a) $\sigma=1.5$ $\mu$s
and $E/\hbar=100\cdot2\pi$ MHz.} \label{fig:twoatomentanglement}
\end{figure}
Figure~\ref{fig:twoatomentanglement}(a) shows the evolution of populations for
realistic experimental parameters, obtained from a numerical propagation of the
Schr\"{o}dinger equation (see caption). The decay due to the finite lifetime of the
$|r\rangle$ states, populated during the process, is incorporated as a decay out of
the system, whereas the $|1\rangle$ and $|2\rangle$ states are treated as long lived
hyperfine sublevels. This implies that if the levels are coupled by single photon
transitions, the field coupling $|1\rangle$ and $|2\rangle$ will have a frequency in
the radiofrequency range, while the laser field coupling $|2\rangle$ and $|r\rangle$
has a wavelength $\approx300$ nm for Cs and Rb. Alternatively the couplings can be
obtained with two-photon transitions involving four optical fields. By appropriately
choosing the laser detunings to compensate for the time-dependent AC Stark shifts
during the pulses, the corresponding five level system of equations with two
intermediate optically excited states can be effectively reduced to the present three
level ladder system.

To investigate the criterion of adiabaticity and the role of the Rydberg interaction
energy $E$, in Fig.~\ref{fig:twoatomentanglement}(b), we show the fidelity of the
creation of the entangled state, $F=|\langle D_2|\psi_f\rangle|^2$, where
$|\psi_f\rangle$ is the final state calculated by propagation of the state vector
with the time-dependent Hamiltonian (\ref{eq:twoatomhamilton}). The simulations show
that, as long as $E$ is sufficiently large to block the population of states with
more than one Rydberg excitation, the exact value of $E$ is not important. With Rabi
frequencies $\Omega_{\textrm{max},1}/2\pi=\Omega_{\textrm{max},r}/2\pi=10$ MHz, a
Rydberg energy shift of $E/\hbar=50\cdot2\pi$ MHz is sufficient. The time window
where $|r\rangle$ is populated is determined by the pulse width and it is desirable
to use the smallest possible width that does not violate adiabaticity, yielding a
total time of entanglement generation of 3-4 $\mu$s, which is short compared with the
radiative lifetime of the highly excited Rydberg state $\gtrsim100$ $\mu$s.

The two-atom entanglement scheme can be modified to create a two-qubit phase gate. To
this end we apply a second STIRAP sequence with phase shifted Rabi frequencies and
the pulses in reversed order, cf. Fig.~\ref{fig:ladderandpulses}(c), that transfers
the population back to the $|11\rangle$. With a non-vanishing relative phase
$\phi_r(t)$ between $\Omega_1$ and $\Omega_r$ the dark state (\ref{eq:twoatomdark})
acquires the geometric phase \cite{berry84}
\begin{equation}\label{eq:geometric11}
\gamma_{2}=-\int\frac{\cos^2\theta\sin^2\theta}{\cos^4\theta+2\sin^4\theta}d\phi_r.\notag
\end{equation}
When we supplement the atomic level scheme with another qubit state $|0\rangle$, which is
uncoupled from the STIRAP pulses, the gate performed by the two STIRAP processes amounts
to multiplication of all two-bit register states by phase factors corresponding to a
controlled two qubit phase gate with phase $\Delta\varphi=\gamma_{2}-2\gamma_{1}$.
$\gamma_{2}$ is only acquired when the pulses overlap, while $\gamma_{1}$ is acquired in
between the two pulse sequences, and $\Delta\varphi$ can, e.g., be controlled by
adjusting $\Delta T$.

We now show that when more than two atoms in $|1\rangle$ are subject to the STIRAP
pulse sequence they also become entangled. Provided all atoms are localized within a
region of a few $\mu$m, the transition towards the Rydberg states is restricted to
the coupling of collective states with either zero or one Rydberg atom and we
implement this by truncating the basis so it only includes states with zero or one
atom in the $|r\rangle$ state and write the symmetric basis states of the system as
$\{|n_1,n_2=N-n_1,0\rangle,|n_1,n_2=N-n_1-1,1\rangle\}$, where $N$ is the total
number of atoms, $n_1$ and $n_2$ the number of atoms in $|1\rangle$ and $|2\rangle$
and $0$ or $1$ indicates if $|r\rangle$ is populated with zero or a single atom. The
interaction with the radiation fields $\Omega_1$ and $\Omega_r$,
$\sum^N_{j=1}-\frac{1}{2}(\Omega_1(t)|2\rangle_j\langle1|+\Omega_r(t)|r\rangle_j\langle2|
+ \textrm{h.c.})$, can now be rewritten
\begin{align}\label{eq:H2}
H(t)&=H_{J_x}(t)+H_{JC}(t),
\end{align}
with variable strengths, representing the coupling by the fields driving the lower and the upper transition, respectively. The dynamics of the
lower levels can be rewritten in a collective spin description, and in the accompanying Schwinger oscillator description,
\begin{equation}
H_{J_x}(t)=-\hbar\Omega_{1}(t)J_x(t)
=-\frac{1}{2}\hbar\Omega_{1}(t)(a_1^{\dag}a_2+a_1a_2^{\dag}),
\end{equation}
where $a_i^{(\dag)}$ are annihilation (creation) operators, with conventional oscillator
commutator relations, for the number of atoms in $|i\rangle$. The upper transition is
described by
\begin{equation}
H_{JC}(t)=-\frac{1}{2}\hbar\Omega_r(t)(a_2\sigma^+ +a_2^{\dag}\sigma^-)
\end{equation}
where we use Pauli matrices, $\sigma^+$ and $\sigma^-$, to describe the transfer of
population between states with zero and one Rydberg excitation in the atomic
ensemble. This Hamiltonian is the quantum optical Jaynes-Cummings (JC) Hamiltonian,
introduced originally to describe the resonant interaction between a two-level atom
and quantized light \cite{cummings65}. JC dynamics has been implemented in strong
coupling cavity QED experiments, where non-classical states of light, such as Fock
states and quantum superposition states, are produced efficiently \cite{auffeves03}.
$H_{JC}$ also describes the motion of laser driven trapped ions where it has been
used to generate various non-classical states \cite{meekhoff96,leibfried96}.
Following these proposals, JC dynamics is sufficient to produce a variety of
interesting states of an atomic ensemble by the Rydberg blockade. Because of its
robustness, however, we shall proceed here, and explore the states resulting from the
STIRAP protocol.

\begin{figure}[htbp]
  \centering
  \includegraphics[width=0.4\textwidth]{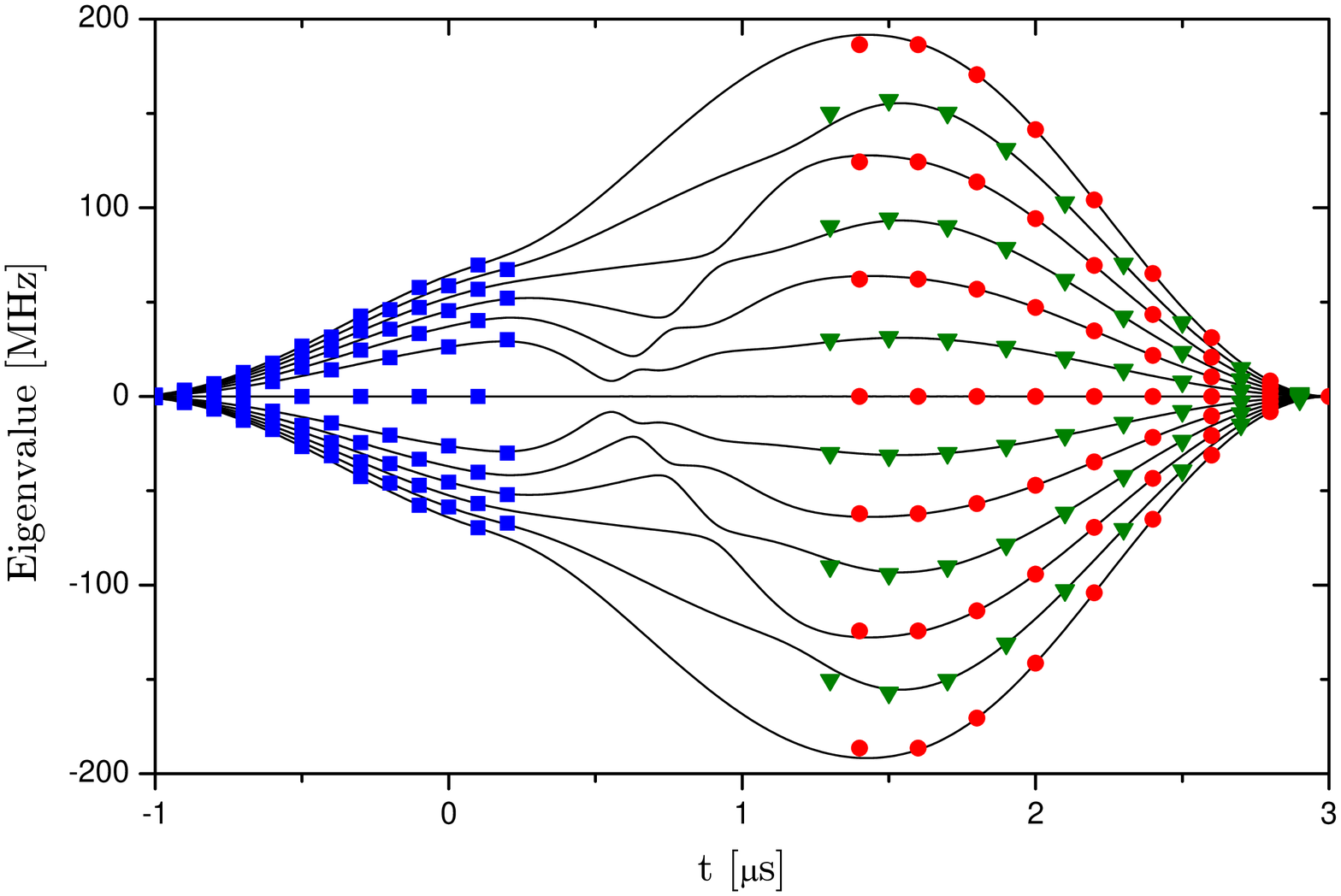}
  \caption{(color online) Eigenvalues of $H(t)/\hbar$ of (\ref{eq:H2}) for $N=6$. Black solid curves show the results of a numerical
  diagonalization of the full Hamiltonian. The squares on the left show the
  analytical eigenvalues of $H_{JC}(t)/\hbar$ and the
  bullets and triangles on the right show the analytical eigenvalues of $H_{J_x}(t)/\hbar$ for zero and one Rydberg excitation, respectively. Parameters used are $\Omega_{max,1}/2\pi=10$ MHz, $\Omega_{max,r}/2\pi=10$
  MHz, $\sigma=1.5$ $\mu$s, $\Delta t=1.1$ $\mu$s}
  \label{fig:eigenvalues}
\end{figure}

The Hamiltonian in (\ref{eq:H2}) has a dark state with zero valued energy eigenvalue
throughout the STIRAP process. This follows because the ``parity'' operator
$\Theta=(-1)^{\hat{n}_2}$, which inverts the sign of the operators $a_2$ and
$a_2^{\dag}$, anti-commutes with $H$, $H\Theta=-\Theta H$. Any eigenstate
$|\psi\rangle$ of $H$ with energy eigenvalue $\mathcal{E}$, then has a partner,
$\Theta|\psi\rangle$ with eigenvalue -$\mathcal{E}$ and the energy eigenvalue
spectrum is symmetric around zero. The number of eigenstates is odd because $N$ atoms
induce $N+1$ different states with no atoms in $|r\rangle$ and $N$ states with one
atom in $|r\rangle$, and hence there must always be a state with eigenvalue zero. The
full curves in Fig.~\ref{fig:eigenvalues} show the $13$ energy eigenvalues for $6$
atoms found from a numerical diagonalization of Eq.~(\ref{eq:H2}), and they clearly
confirm the existence of the dark state throughout the pulse sequence.

The STIRAP process starts with $\Omega_1=0$ and hence initially
$H(t)=H_{JC}(t)=-\frac{1}{2}\hbar\Omega_r(t)(a_2\sigma^++a_2^{\dag}\sigma^-)$ with
the eigenvalues $\{0,\pm\frac{1}{2}\hbar\Omega_r(t)\sqrt{n_2+1}\}$, as shown with the
squares in Fig.~\ref{fig:eigenvalues}. The dark state is $|n_1=N,n_2=0,n_r=0\rangle$
with all atoms in state $|1\rangle$. Adiabaticity ensures that we remain in the dark
state of the full Hamiltonian and when $\Omega_1$ is turned on and $\Omega_r$
reduced, the system ends up in the dark state of
$H(t)=H_{J_x}(t)=-\hbar\Omega_1(t)J_x$. In this state, the system can have either no
or a single Rydberg excitation leaving $K=N$ or $K=N-1$ atoms in the $|1\rangle$ and
$|2\rangle$ states. The eigenvalues of $-\Omega_1(t) J_x$ are
$-\Omega_1(t)\{-K/2,-K/2+1,...,K/2-1,K/2\}$ and $\mathcal{E}=0$ occurs for $K$ even.
Figure~\ref{fig:eigenvalues} shows the eigenvalues of $-\Omega_1(t)J_x$ when
$|r\rangle$ is not populated (bullets) and when one atom is excited to $|r\rangle$
(triangles) for $N=6$. For $N$ even, the dark state does not populate $|r\rangle$,
while for $N$ odd, the final dark state is the state with one Rydberg excitation and
$N-1$ atoms in the $J_x=0$ eigenstate. In general we write the final dark state
\begin{equation}
|D_N\rangle=\begin{cases}
|J_x=0\rangle  & \text{if $N$ is even}\\
(|J_x=0\rangle\otimes|r\rangle)_{sym} & \text{if $N$ is odd}
\end{cases},\label{eq:finalstate}
\end{equation}
where $(.)_{sym}$ indicates that the state is symmetrized with respect to $|r\rangle$,
such that any atom is Rydberg excited with equal weight.

The STIRAP protocol produces a $|J_x=0\rangle$ multi-particle entangled state, and
precisely this state reaches the Heisenberg limit of phase sensitivity in entanglement
enabled precision metrology \cite{Holland93, Bouyer97}. If the two lower states are,
e.g., the hyperfine states of the Cs clock transition, the presented entanglement scheme
thus constitutes an ideal preparation of the system for an atomic clock.

The state (\ref{eq:finalstate}) has none or a single Rydberg excited atom, depending
on the number of atoms initially in the $|1\rangle$ state, $n_1$, being even or odd.
Following \cite{yurke95}, this can be used to prepare a Greenberger-Horne-Zeilinger
(GHZ) state \cite{greenberger89} of the system. We first prepare all our atoms in a
spin coherent state, i.e., a product of superpositions of two ground states
$|0\rangle$ and $|1\rangle$,
\begin{equation}
\left(\frac{|0\rangle+|1\rangle}{\sqrt{2}}\right)^{\otimes
N}=\sum_{n_1=0}^{N}\sqrt{\left(\begin{array}{c}N
\\n_1\end{array}\right)}\left(\frac{1}{\sqrt{2}}\right)^{N}|n_0,n_1\rangle
\end{equation}
where $n_1$ is the number of atoms in the state $|1\rangle$ and $n_0=N-n_1$. Applying
the pair of STIRAP processes in Fig.~\ref{fig:ladderandpulses}(c) to the entire
system, transfers each component $|n_0,n_1\rangle$ to a state with none or a single
Rydberg excitation and back. An energy shift of the Rydberg state or a phase shift of
the laser coupling the Rydberg state can now be used to provide states populating the
Rydberg state with a phase, $i=e^{i\pi/2}$.  We can write this phase,
$e^{i\eta({n_1})}=(e^{i\pi/4}+(-1)^{n_1}e^{-i\pi/4})/\sqrt{2}$ leading after the
second STIRAP process to the final GHZ state,
\begin{align}\label{eq:ghzstate}
&|\psi_f\rangle=\sum_{n_1=0}^{N}\sqrt{\left(\begin{array}{c}N
\\n_1\end{array}\right)}\left(\frac{1}{\sqrt{2}}\right)^{N}e^{i\eta({n_1})}|n_0,n_1\rangle\\\notag
&=\frac{e^{i\pi/4}}{\sqrt{2}}\left(\frac{|0\rangle+|1\rangle}{\sqrt{2}}\right)^{\otimes{N}}+
\frac{e^{-i\pi/4}}{\sqrt{2}}\left(\frac{|0\rangle-|1\rangle}{\sqrt{2}}\right)^{\otimes{N}}.
\end{align}

When $N$ is increased, to  maintain adiabaticity one must require longer pulses
and/or a stronger Rydberg blockade. As indicated by the energy spectra in
Fig.~\ref{fig:eigenvalues}, there are crossing regions, where special care should be
taken, and we anticipate that control theory may be used to find optimal pulse
shapes. Using the simple $\sin^2$-pulses, we have investigated the preparation of the
$|J_x=0\rangle$ (\ref{eq:finalstate}) and GHZ (\ref{eq:ghzstate}) states for up to
$N=10$ taking into account the coupling to states with two Rydberg excited atoms and
the accompanying energy shift. Note that the explicit Rydberg-Rydberg interaction
energy does not anti-commute with $(-1)^{{\hat{n}}_2}$, and the spectrum is not
symmetric around zero to this order of precision - the non-resonant coupling to the
states with the large interaction energy shift causes a perturbation of the energy
levels in the figure. This perturbation implies a small dynamic phase on the dark
state, even if adiabaticity is not violated. We find a population of the
$|J_x=0\rangle$ or GHZ state above $0.995$ for peak Rabi frequencies of $10\cdot2\pi$
MHz, a Rydberg interaction, $E/2\pi=400$ MHz and pulse widths of $50$ $\mu$s. These
numbers require the use of Rydberg levels with $n\geq100$ where a lifetime above $1$
ms can be achieved for atoms in a cryogenic environment \cite{Saffman05}. If higher
Rabi frequencies can be achieved, adiabaticity can be maintained with shorter pulses
and hence Rydberg states with shorter lifetimes can be used.

In conclusion, we have demonstrated that the Rydberg excitation blockade mechanism in
conjunction with rapid adiabatic passage processes provide rich opportunities to
prepare two- and multi-atom entangled states with confined samples of atoms, e.g., in
optical dipole traps or small lattice arrays. Our calculations indicate the
possibility of high fidelity generation of entangled states and quantum
superpositions states with tens of atoms, and we propose to apply control theory
methods to optimize pulse shapes and reach even larger systems.

\begin{acknowledgments}
This work is supported by the Danish Research Agency (Grant. No. 2117-05-0081) and by
the ARO-DTO (Grant. No. 47949PHQC). LBM acknowledges the hospitality of the Physics
Department at the University of Otago, New Zealand, where part of this work was
performed.
\end{acknowledgments}

\end{document}